\documentclass[letterpaper,twocolumn]{jpsj3}
\usepackage{txfonts}

\title{Negatively Charged Muonium and Related Centers in Solids}

\author{Takashi~U.~Ito$^1$\thanks{ito.takashi15@jaea.go.jp}, Wataru~Higemoto$^{1,2}$, and Koichiro~Shimomura$^3$}
\inst{$^1$Advanced Science Research Center, Japan Atomic Energy
 Agency, Tokai, Ibaraki 319-1195, Japan \\
$^2$Department of Physics, Tokyo Institute of Technology, 
Meguro, Tokyo 152-8551, Japan \\
$^3$Institute of Materials Structure Science, High Energy
Accelerator Research Organization (KEK), Tsukuba,
Ibaraki 305-0801, Japan} 

\abst{Muonium (Mu) centers formed upon implantation of $\mu^+$ 
in a solid have long been investigated as an experimentally accessible model of isolated
hydrogen impurities.
Recent discoveries of hydridic centers formed at oxygen vacancies have
stimulated a renewed interest in H$^-$ and corresponding Mu$^-$ centers in oxides.
However, the two diamagnetic centers, Mu$^+$ and Mu$^-$, are difficult to
separate spectroscopically.
In this review article, we summarize established and
developing methodologies for identifying Mu$^-$ centers in solids, and review
recent Mu studies on said centers supposedly formed in mayenite, oxygen-deficient
SrTiO$_{3-x}$, and BaTiO$_{3-x}$H$_x$ oxyhydride.}

\begin{document}
\newcommand{\bi}[1]{\ensuremath{\boldsymbol{#1}}} 

\maketitle

 \section{Introduction}
 Hydrogen is a ubiquitous impurity in most semiconductors and
 insulators, which can unintentionally enter crystalline lattices during
 crystal growth as well as subsequent processing.
 The incorporated hydrogen often has a significant impact on structural
 and electrical properties of these materials in spite of low
 hydrogen solubilities.\cite{cox09}
 It is beneficial to application of Si, where hydrogen passivates dangling bonds of 
 undercoordinated Si at vacancies, impurities, surfaces, and interfaces.
 Meanwhile, hydrogen can also passivate deliberate dopants by forming
 H-dopant complexes.
 It has also been recognized that hydrogen isolated from other defects
  has an electrical activity, involving all allowed charge 
 states: H$^+$, H$^0$, and H$^-$. The isolated hydrogen often shows
 amphoteric behavior, acting as both deep donor and acceptor in
 accordance with the Fermi level position in the band gap.\cite{walle03}
 It can also behave as shallow
 donor, which has been observed in several oxides, such as
 ZnO,\cite{hofmann02} or as shallow acceptor.
 Because of these complex electrical activities, a detailed understanding of
 hydrogen and related defects is vital for application of host materials.
 However, it is very difficult to experimentally characterize hydrogen
 in such trace quantities, particularly isolated hydrogen.

 Muonium (Mu) centers formed upon implantation of $\mu^+$ in a solid 
 have long been used as an experimentally accessible model of 
 hydrogen.\cite{cox09}
Indeed, the electronic states of H and Mu centers are nearly identical, 
as can be seen when comparing atomic Mu$^0$($=\mu^+e^-$) and H$^0$ in a 
 vacuum: they have almost the same reduced mass.
Microscopic insight into Mu centers can be obtained using the muon spin
 rotation, relaxation, and resonance ($\mu^+$SR) spectroscopy, which is
 analogous to the $^1$H nuclear magnetic resonance (NMR) spectroscopy.
 Technical details of the $\mu^+$SR method have been previously outlined
 in existing literature, such as Ref.~[\citen{yaounc10}].
 Since only a small number of muons stay in a sample at a time,
 the $\mu^+$SR method can provide information on isolated Mu
 centers. In principle, the electronic structures of Mu centers can be
  investigated more precisely than those of corresponding H centers by
 measuring hyperfine frequencies or chemical shifts because the gyromagnetic ratio
 for $\mu^+$ is roughly three times larger than that for
 $^1$H.
  It should be also emphasized that the Mu centers observed may not be in 
 global equilibrium\cite{lichti08} because the formation of Mu centers and the
 observation of their states are completed in a relatively short time
 scale, which is comparable to the muon lifetime $\sim 2.2~\mu$s.
Therefore, the $\mu^+$SR spectroscopy is potentially useful for
 investigating excited hydrogen configurations in heavily hydrogenated
 compounds as well, where Mu species are expected to behave as
 metastable excess hydrogen and interact with preexisting hydrogen in
 the host lattice\cite{ito17}.

\begin{figure}
\begin{center}
\includegraphics[scale=0.60]{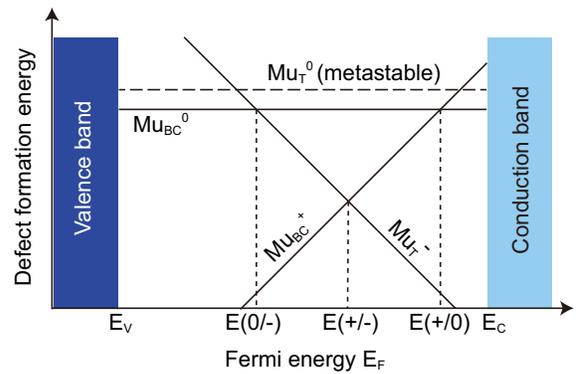}
\caption{(Color online) Defect-formation energy diagram for interstitial Mu in GaAs.}
 \label{GaAs_energy_diagram}
\end{center}
\end{figure}
 
Historically, the Mu approach has played an important role in
microscopically establishing the amphoteric behavior of isolated interstitial
hydrogen in many semiconductors, such as Si and GaAs.
Figure~\ref{GaAs_energy_diagram} shows the Mu defect-formation energy
diagram for GaAs obtained from comparisons with a theoretical model
of H\cite{walle03} and the experimental results for Mu.\cite{lichti08} This
indicates that the interstitial Mu counteracts prevailing conductivity as
a donor Mu$^+$ or an acceptor Mu$^-$. In addition, the dominant equilibrium charge
state changes from positive to negative at the charge-transition level $E(+/-)$
as a function of the Fermi energy $E_F$.
According to Van de Walle and Neugebauer, $E(+/-)$ for H 
should be universally pinned at a specific energy below the
vacuum level.\cite{walle03} This was experimentally confirmed
by Lichti {\it et al.} for Mu analogs in Si, Ge, GaAs, GaP, ZnSe, and
6H-SiC.\cite{lichti08}
The defect-formation energy diagram also suggests that the dominant
equilibrium charge state is always Mu$^+$ (Mu$^-$) when $E(+/-)$ is in
the conduction (valence) band.
Indeed, such shallow donor (acceptor) behavior has been theoretically predicted for interstitial H in some
oxides, nitrides, and
sulfides,\cite{walle00,limpijumnong01,kilic02,iwazaki10,iwazaki14,varley13}
as well as indirectly confirmed by observing shallow donor (acceptor) Mu states.\cite{gil99,shimomura02,davis03,shimomura04,cox06a,cox06b,ito13,shimomura16,carroll14}

Recent discoveries of hydridic centers formed at oxygen vacancies
(V$_{\rm O}$) have stimulated a renewed interest in H$^-$ and corresponding Mu$^-$ centers
in oxides. Theoretically, the most stable form of hydrogen trapped at
V$_{\rm O}$ is H$^-$, where doubly charged V$^{2+}_{\rm
O}$ changes into singly charged H$_{\rm O}^+$.\cite{liu17,iwazaki10,iwazaki14,zhang14}
This indicates that the substitutional H$^-$ serves as a single donor in
sharp contrast to the acceptor behavior of the interstitial H$^-$.
A large amount of O$^{2-}$ in the host lattice can be replaced with
H$^-$ due to its structural stability (up to $\sim$ 40\% in
SmFeAsO$_{1-x}$H$_x$\cite{hanna11} and $\sim$ 20\%
in BaTiO$_{3-x}$H$_x$\cite{kobayashi12}).
The substitutional H$^-$ is becoming more and more important 
as an electron dopant in oxides\cite{hanna11} and a novel charge carrier
for solid-state ionic conductors.\cite{kobayashi12}
However, microscopic insight into substitutional H$^-$
and related defect complexes is currently limited.

The $\mu^+$SR spectroscopy is potentially useful for the study of Mu$^-$
analog in oxides, where attention should be focused on distinguishing
between Mu$^-$ and Mu$^+$.
Generally, it is difficult to separate these diamagnetic Mu species 
spectroscopically because they have quasi-identical responses to the 
applied magnetic field $B$.
In this review article, we summarize new and existing methodologies for identifying
Mu$^-$ centers.
In addition, we review the recent $\mu^+$SR studies on said centers supposedly formed in
mayenite, oxygen-deficient SrTiO$_{3-x}$, and BaTiO$_{3-x}$H$_x$ oxyhydride
upon implantation of $\mu^+$.
Finally, we provide concluding remarks on the future prospects of Mu$^-$ studies.


\section{Methodologies for Identifying Mu$^-$ and Related Centers}
\label{method}
\subsection{Local Structures}
\label{method_structure}
\begin{figure}
\begin{center}
\includegraphics[scale=0.65]{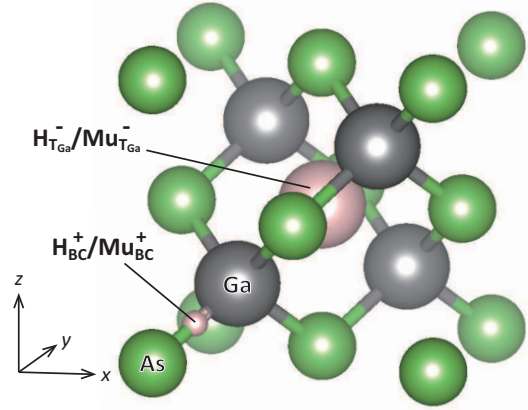}
 \caption{(Color online) Correlations between the sites and favorable charge states of diamagnetic H/Mu centers in GaAs.\cite{estreicher95} The structure was drawn using VESTA.\cite{vesta3}}
 \label{GaAs-BC-T}
\end{center}
\end{figure}

Computational studies on H impurities in semiconductors have revealed
that correlations exist between their crystallographic sites
and favorable charge states (Fig.~\ref{GaAs-BC-T}).\cite{estreicher95}
However, it is difficult to identify the structures of H-related defects 
due to their random distribution as well as low concentrations.
Therefore, the information on the structures of Mu centers obtained from 
$\mu^+$SR spectroscopy is vital with respect to identifying Mu/H charge states.
The structures of paramagnetic Mu$^0$ centers can be easily
estimated from hyperfine anisotropies.\cite{kiefl85,schneider93}
This is also the case for shallow and polaronic Mu centers, where a 
Mu$^+$ or Mu$^-$ core weakly binds an unpaired electron or hole.\cite{gil99,shimomura02,davis03,shimomura04,cox06a,cox06b,ito13,shimomura16,ito19,carroll14}
Contrastingly, diamagnetic Mu$^{\pm}$ centers are much more
difficult to characterize because they do not have 
unpaired electrons, and, accordingly, they lack hyperfine interactions.
In such cases, magnetic-dipolar and electric-quadrupolar interactions
between a muon and surrounding nuclei can be used to identify the structures of
diamagnetic Mu centers. These interactions can be efficiently studied using the muon level-crossing
resonance ($\mu$LCR) technique,\cite{cox09} as demonstrated in heavily doped
GaAs.\cite{chow95,schultz05}

Two paramagnetic and two diamagnetic Mu centers are primarily observed
in GaAs.\cite{cox09} In heavily doped samples, only the diamagnetic Mu centers are detectable.
Theoretical studies on H impurities\cite{estreicher95} suggest that
they should be assigned to Mu$^-$ at the tetrahedral site surrounded by
Ga atoms (T$_{\rm Ga}$) or Mu$^+$ near the bond-center (BC)
position (Fig.~\ref{GaAs-BC-T}). The Mu$^-_{\rm T}$ and Mu$^+_{\rm BC}$ centers are
theoretically stable in $n$-type and $p$-type carrier-rich environments, respectively.

Chow {\it et al.} investigated the diamagnetic Mu center
formed in $n$-type GaAs (Si concentrations between $2.5\times 10^{18}$
and $5 \times 10^{18}$ cm$^{-3}$) using $\mu$LCR and conventional transverse-field
(TF) $\mu^+$SR techniques to establish the Mu$^-_{\rm T}$ state from a structural point
of view.\cite{chow95}
Herein, the time-integrated muon spin polarization $\bar{P}$ was
obtained in the longitudinal field (LF) configuration\cite{yaounc10} as a function of $B$.
The spin Hamiltonian relevant to this experiment comprises the Zeeman
interactions for the muon and surrounding nuclei, the magnetic-dipolar
interactions between the muon and the nuclei, and the electric-quadrupolar
interactions for the nuclei that have an electric field gradient (EFG)
primarily caused by the diamagnetic Mu.
Resonant cross relaxation between the
muon and the nuclei, observed as dips in $\bar{P}(B)$, occurs 
when the muon Zeeman splitting matches the separation of the
combined quadrupolar and Zeeman energy levels for the nuclei.
The resonance position depends on the electric quadrupolar parameter $Q_i$ for 
the on-resonance nucleus $i$ and the angle $\theta_i$ between \bi{B} and
the muon-nucleus axis. For a given $Q_i$ and $\theta_i$, the intensity 
of the resonance line is a function of the muon-nuclear
dipolar coupling $D_i$, which is dependent on the muon-nucleus distance $r_i$.
Chow {\it et al.} observed a couple of doublets in the
$\bar{P}$ spectrum under \bi{B}~$||~\langle 001 \rangle$, which were
assigned to the two spin-3/2 isotopes of $^{69}$Ga and
$^{71}$Ga. Another resonance was identified at the low-field end of the
spectrum, assigned to $^{75}$As with spin 3/2.
The small number of lines and the structure of the Ga doublets indicate
that the principal axis of the EFG tensor, or the muon-Ga axis, is
parallel to $\langle 111 \rangle$ at $\theta_{\rm Ga}\sim 54.7^{\circ}$,
which is compatible with the BC, T$_{\rm Ga}$, and AB$_{\rm Ga}$
(antibonding to a Ga atom) sites. The $r_i$ parameter was estimated as
$r_{\rm Ga}=2.199(7)$~\AA~ and $r_{\rm As}=2.72(5)$~\AA~ from a combined
analysis of the $\mu$LCR spectrum and the TF-$\mu^+$SR Gaussian width as
functions of \bi{B}.
The results are consistent with the Mu$^-$ center at the T$_{\rm Ga}$
site, the $r_{\rm Ga}$ of which is roughly 10\% shorter than expected for
an undistorted GaAs lattice.



\begin{figure}
\begin{center}
\includegraphics[scale=0.75]{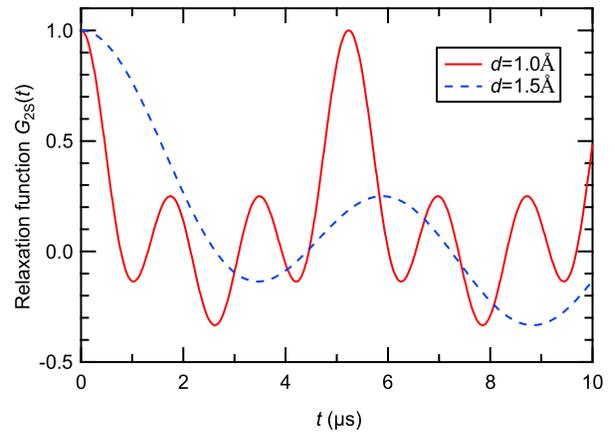}
\caption{(Color online) ZF muon spin relaxation function $G_{\rm 2S}(t)$ for the 2S
 state with $\gamma_{I}/2\pi=42.58$~MHz/T (muon-$^1$H complex).
 The solid and dashed curves correspond to the configurations with 
 $d=1.0$~\AA~ and 1.5~\AA, respectively.}
 \label{G2S}
\end{center}
\end{figure}

The $\mu$LCR technique cannot be applied to systems that do not contain
nuclei with a finite nuclear quadrupolar moment. Even in such cases, the
local structures of diamagnetic Mu centers are occasionally determined 
via the magnetic-dipolar interactions between a muon and
the surrounding nuclei. Characteristic oscillating features appear in
zero field (ZF) $\mu^+$SR spectra when the muon is dipole coupled to a
small number of nuclei, thus creating an entangled a-few spin
system.\cite{brewer86,nishiyama03,lancaster07,kadono08,ito09,sugiyama10}
Such a state often forms in materials containing
fluorine\cite{brewer86,nishiyama03,lancaster07} or
hydrogen\cite{kadono08,ito09,sugiyama10}
with relatively strong electronegativity, which causes muons to
preferentially localize in their vicinity.
In addition, the $^{19}$F and $^1$H nuclei with spin 1/2 have relatively large
dipolar moments.
Here we consider an entangled spin system consisting of a muon and a
spin-1/2 nucleus.
The powder-averaged muon spin relaxation function for the two-spin state
(hereafter, referred to as 2S) in ZF can be expressed as
\begin{align}
G_{\rm 2S}(t)&=\frac{1}{6}+\frac{1}{6}\cos(2\pi f_d t)\nonumber\\
&+\frac{1}{3}\cos(\pi f_d t)+\frac{1}{3}\cos(3\pi f_d t),\label{eq_2S}\\
&f_d=\frac{\mu_0\hbar\gamma_{\mu}\gamma_{I}}{8\pi^2 d^3},\label{eq_fd}
\end{align}
where $d$ is the distance between $\mu^+$ and the spin-1/2 nucleus 
and $\gamma_{\mu}$ and $\gamma_{I}$ are the gyromagnetic ratios for 
$\mu^+$ ($\gamma_{\mu}/2\pi=135.53~{\rm MHz/T}$) and the nucleus
($\gamma_{I}/2\pi=42.58~{\rm MHz/T}$ for $^1$H or $40.08~{\rm MHz/T}$
for $^{19}$F), respectively (Fig.~\ref{G2S}).\cite{nishiyama03,stone05}
The muon-nucleus distance, $d$, obtained by fitting eq.~(\ref{eq_2S}) is
useful with respect to estimating the local structure of a diamagnetic Mu center.
This method was applied to identify an H$^-$-Mu$^-$ complex supposedly
formed in the BaTiO$_{3-x}$H$_x$ oxyhydride upon implantation of $\mu^+$
by Ito {\it et al.}\cite{ito17}, as reviewed in Section~\ref{review_btoh}.

\subsection{Thermal Properties}
\label{method_thermal}
As the temperature increases, the Mu site changes and/or charge-state
transitions are thermally activated.
Characteristic energies relevant to these processes are obtained from the
temperature dependences of diamagnetic and paramagnetic Mu amplitudes
and muon spin relaxation rates.\cite{cox09}
The Mu species involved can be identified by mapping these energies to
theoretically allowed transitions.
The H/Mu defect-formation energy diagram, such as
Fig.~\ref{GaAs_energy_diagram}, obtained from first-principles
calculations provides potential candidates for the transitions that are linked
to Mu sites and charge states.
In low-carrier systems, final Mu states can be different from
promptly formed Mu states, which may be reached via a delayed process with a time
constant comparable to the muon lifetime $\sim 2.2~\mu$s.
The radio frequency (RF) $\mu^+$SR technique is suitable for
investigating such final states, as demonstrated in the studies of Mu
centers in Si, Ge, and GaAs.\cite{kreitzman91,hitti99,lichti99,lichti07,lichti08}


Longitudinal $T_1$ relaxation is often linked to 
cyclic charge-state transitions between Mu$^0$ and Mu$^-$ or Mu$^+$ in thermal equilibrium.
Chow {\it et al.} established that the Mu defect in heavily doped
$n$-type GaAs:Si serves as a deep recombination center via the
observation of a Mu$^{-/0}_{\rm T}$ charge cycle.\cite{chow96} The
longitudinal relaxation rate 1/$T_1$ associated with the Mu$^{-/0}_{\rm
T}$ dynamics is expressed as a function of $\lambda_{-0}$, $\lambda_{-0}$,
$A_{\mu}$, and $B$,
where $\lambda_{-0}$ and $\lambda_{0-}$ are the rates of conversion for
${\rm Mu}_{\rm T}^-\rightarrow{\rm Mu}_{\rm T}^0$ and ${\rm Mu}_{\rm
T}^0\rightarrow{\rm Mu}_{\rm T}^-$,
respectively, and $A_{\mu}$ is the hyperfine coupling constant for
Mu$^0_{\rm T}$. The $\lambda_{-0}$ (hole capture rate) exhibits 
Arrhenius-like behavior with an activation energy of 1.66(7)~eV.
This value is approximal to that of the band gap in GaAs, suggesting 
that a band-gap excitation governs this process.
Contrastingly, the $\lambda_{0-}$ (electron capture rate) is 
independent of temperature and much larger than the $\lambda_{-0}$. 
This is expected since the electron-carrier concentration is nearly
temperature-independent in heavily doped $n$-type GaAs.

The $T_1$ relaxation can also occur when a Mu$^{+/0}$ charge cycle is
activated. The values and behavior of
charge-transition rates associated with this process in intrinsic Si are considerably 
different from those of $\lambda_{-0}$ and $\lambda_{0-}$ in heavily
doped $n$-type GaAs.\cite{chow93} 
This suggests that the charge state of diamagnetic Mu species involved
in a Mu-charge cycle can be distinguished by carefully investigating the 
charge-transition rates.

\subsection{Photo-Detachment of the Second Electron from Mu$^-$}
\label{method_laser}
The optical excitation of electrons trapped at point defects to the
conduction band (CB) is a useful method of investigating the corresponding defect
levels formed in the band gap. Such a technique, combined with
$\mu^+$SR, is also helpful with respect to distinguishing between Mu$^-$ and Mu$^+$
in semiconductors.
The photoexcitation of the second electron from Mu$^-$ to the CB causes
a paramagnetic Mu$^0$ center, which can be described as
\begin{equation}
{\rm Mu}^- + h\nu \rightarrow {\rm Mu}^0+e^- {\rm (CB)},
\end{equation}
where $h\nu$ denotes a photon with energy corresponding to the optical
excitation of the second electron to the CB minimum.
The photogenerated Mu$^0$ center is detectable using conventional
$\mu^+$SR techniques due to the strong hyperfine interactions.

Shimomura {\it et al.} conducted optical $\mu^+$SR experiments to
identify the Mu$^-_{\rm T}$ acceptor in $n$-type GaAs using a high-power
pulse laser.\cite{shimomura10,shimomura12} The experiments were
performed at the port 2 of the RIKEN-RAL muon facility, UK, using a
broadband OPO laser system pumped by a 355-nm beam from a Nd:YAG laser.
The laser system was operated in pulse mode at a 25-Hz repetition rate
and synchronized to the arrival of muon pulses to a single-crystalline
GaAs wafer doped with $3\times10^{16}$-cm$^{-3}$ Si. The decrease in 
$\mu^+$SR asymmetry associated with the Mu$^-_{\rm T} \rightarrow {\rm
Mu}^0_{\rm T}$ conversion was recorded as a function of photon energy, which
was scanned from 0.8 to 1.5~eV.
Shimomura {\it et al.} found a broad feature centered at around 1~eV
after normalizing the asymmetry decrease by laser power. This is 
consistent with the results obtained from the temperature dependence of 
RF-$\mu^+$SR amplitude assigned for the Mu$_{\rm T}^-\rightarrow {\rm Mu}_{\rm
T}^0+e^-{\rm (CB)}$ excitation.\cite{lichti07} Moreover, a sharp
feature was identified at around 1.5~eV, which was attributed to the spin or charge scattering
between Mu$_{\rm T}^-$ and photoexcited electrons. Subsequent experiments
using circularly polarized laser light revealed that this effect depends
on the polarization direction of conduction electrons with respect to the muon
polarization direction; however, the exact mechanism is yet to be
clarified.\cite{yokoyama12, gu14}

\subsection{Chemical Shifts}
\label{method_chemical_shift}
Recent systematic $^1$H-NMR studies in Ca and Sr-mayenites (C12A7 and
S12A7) have revealed that $^1$H chemical shifts are useful with respect to 
distinguishing nominally H$^+$ and H$^-$ species.\cite{hayashi14}
Mayenites comprise a class of cage-structured compounds that contain
``extraframework'' anions in their cages, such as O$^{2-}$, OH$^-$,
H$^-$, and $e^-$.\cite{salasin17}
Hayashi {\it et al.} prepared mayenite samples incorporating OH$^-$ or
H$^-$ extraframework anions with concentrations approximal to theoretical
maxima and measured the isotropic chemical shifts $\delta_{iso}$
for said H species using the magic-angle-spinning technique.
They obtained a $\delta_{iso}$ of $+5.1$~ppm (+6.1~ppm) for H$^-$ 
and $-0.8$~ppm ($-1.3$~ppm) for OH$^-$ (formally H$^+$) in C12A7 (S12A7)
with respect to the tetramethylsilane (TMS) reference.
Surprisingly, the observed values of $\delta_{iso}$ for H$^-$ are larger
than those for H$^+$, falling within the typical range for the H$^+$
state from +20 to 0~ppm.\cite{akitt92}
This result suggests that the electron density around the $^1$H nuclei, 
which is associated with the chemical-shielding effect, is larger for H$^+$ than
H$^-$ species in mayenites.
 First-principles calculations using the periodic
and embedded cluster approaches successfully reproduced such electronic
states as well as the $\delta_{iso}$(H$^-$)~$>$~$\delta_{iso}$(H$^+$)
relation. The experimental and theoretical results for 
$\delta_{iso}$s are consistent with each other within 1~ppm, demonstrating that the combined approach
based on $^1$H chemical shift measurements and first-principles
calculations is useful with respect to 
identifying the charge state of H species.
Systematic surveying of the relation between the $^1$H chemical shift and
the local structure around H species has also been conducted for many oxides,
ionic hydrides, and mixed-anion hydrides. The research suggests that the 
distance between the H species and the coordinating atoms strongly affects 
the chemical shift and that $\delta_{iso}$ for both H$^+$ and H$^-$ is distributed in
a similar range. This indicates that the H$^{\pm}$ species can be
separated by carefully analyzing the chemical shift; however, it 
does not serve as an {\it easy} fingerprint of the nominal charge states.

The chemical shift approach has also been applied to identify Mu$^{\pm}$ species
supposedly formed in C12A7:O$^{2-}$ and C12A7:$e^-$ upon
implantation of $\mu^+$ by Hiraishi {\it et al.},\cite{hiraishi16}
as reviewed in Section~\ref{review_mayenite}.


\section{Recent Mu$^-$ Studies in Oxides}
\subsection{Chemical Shifts of Diamagnetic Mu Centers in
  Mayenite [\citen{hiraishi16}]}
 \label{review_mayenite}
The extraframework H species in C12A7 mayenite have attracted much
attention in association with
persistent photoconductivity in C12A7:H$^-$\cite{hayashi02} and highly
efficient ammonia synthesis using Ru-loaded
C12A7:$e^-$.\cite{kitano12,kitano15} 
Hiraishi {\it et al.} adopted the $\mu^+$SR technique to investigate the
electronic structure and chemical activity of these centers using Mu
analogs.\cite{hiraishi16} 
Here, we focus on their attempts to distinguish extraframework OMu$^-$ (formally Mu$^+$)
and Mu$^-$ species by measuring the Mu chemical shift $K_{\mu}$
 according to the methodology outlined in Section~\ref{method_chemical_shift}.
 These species are supposedly formed in C12A7:O$^{2-}$ and C12A7:$e^-$,
respectively, via
\begin{equation}
\{\rm{O}^{2-}\}+\rm{Mu}^0 \rightarrow \{\rm{OMu}^-\}+\{{\it e}^-\},
\end{equation}
\begin{equation}
\{e^{-}\}+\rm{Mu}^0 \rightarrow \{\rm{Mu}^-\},
\end{equation}
where \{{\it X}\} indicates an extraframework anion $X$ in a cage.

Mu chemical shift measurements were conducted on single-crystalline
samples of C12A7:O$^{2-}$ (pristine insulator) and C12A7:$e^-$
(electride, $n\sim10^{21}$~cm$^{-3}$) in the high-TF 
configuration\cite{yaounc10} at TRIUMF, Canada, using a spin-polarized surface muon
beam. A TF of 6~T was applied along the muon incident axis, which was
nominally perpendicular to the (001) plane.
For both samples, the value of $K_{\mu}$ was mostly independent of temperature
below 300~K. 
The temperature-averaged $K_{\mu}$ was $+0.3(4)$~ppm for C12A7:O$^{2-}$ and $+6.6(4)$~ppm for
C12A7:$e^-$ using CaCO$_3$ as a zero-shift reference.
These results are consistent with corresponding $^1$H chemical shifts,
specifically $\delta_{iso}=-0.8$~ppm for C12A7:OH$^-$ and $\delta_{iso}=+5.1$~ppm for
C12A7:H$^-$, using TMS as a reference.\cite{hayashi14} On the basis of this similarity,
Hiraishi {\it et al.} assigned the chemical state of diamagnetic muons
in the C12A7:O$^{2-}$ and C12A7:$e^-$ samples to \{OMu$^-$\} and
\{Mu$^-$\}, respectively.
However, there seems to be some ambiguity associated 
with the difference in the gyromagnetic ratios for $\mu^+$ and $^1$H 
($\gamma_{\mu}/\gamma_{\rm H}\sim3.2$). Indeed, the value of $K_{\mu}$ for \{Mu$^-$\}
in C12A7:$e^-$ is considerably smaller than expected from the $^1$H
chemical shift for C12A7:H$^-$ ($5.1\times 3.2=16$~ppm). This
discrepancy was tentatively attributed to the metallic environment of
Mu$^-$ in the C12A7:$e^-$ sample ($n\sim 10^{21}$~cm$^{-3}$).\cite{hiraishi16}

\subsection{Possible Charge-State Dynamics of Mu in Oxygen-Deficient
  SrTiO$_{3-x}$ [\citen{shimomura14}]}
 \label{review_odef_sto}
Oxygen vacancies can act as electron donors in SrTiO$_3$, causing a
wide range of interesting phenomena, such as superconductivity\cite{schooley64} and
ferromagnetism.\cite{rice14} The oxygen-deficient SrTiO$_{3-x}$ can be obtained by
annealing the parent band insulator in highly reducing
atmospheres. Hydrogen is often used as a reducing agent in this
process. Systematic annealing studies have revealed that such a treatment does
not only remove oxygen to obtain metallic SrTiO$_{3-x}$,
but that it can also create other types of defects involving
hydrogen.\cite{jalan08} The most common H-related defect in
perovskite oxides is an interstitial H$^+$ bound to an O$^{2-}$ in the
host lattice, which serves as a shallow donor in
SrTiO$_3$\cite{nakayama18}. This behavior has also been confirmed from
Mu viewpoints\cite{salman14,ito19}. Contrastingly, subsequent hydrogen
annealing of the metallic SrTiO$_{3-x}$ causes a decrease in conductivity.\cite{jalan08} 
This contradictory behavior against the interstitial H$^+$ implies that 
the site and charge state of hydrogen introduced by the subsequent
annealing is considerably different from those of the interstitial H$^+$.
First-principles studies suggest that the most stable form of hydrogen
in SrTiO$_{3-x}$ 
is H$^-$ located at the anion site, where doubly charged V$^{2+}_{\rm
O}$ changes into singly charged H$_{\rm O}^+$.\cite{iwazaki14}

\begin{figure}
\begin{center}
\includegraphics[scale=0.7]{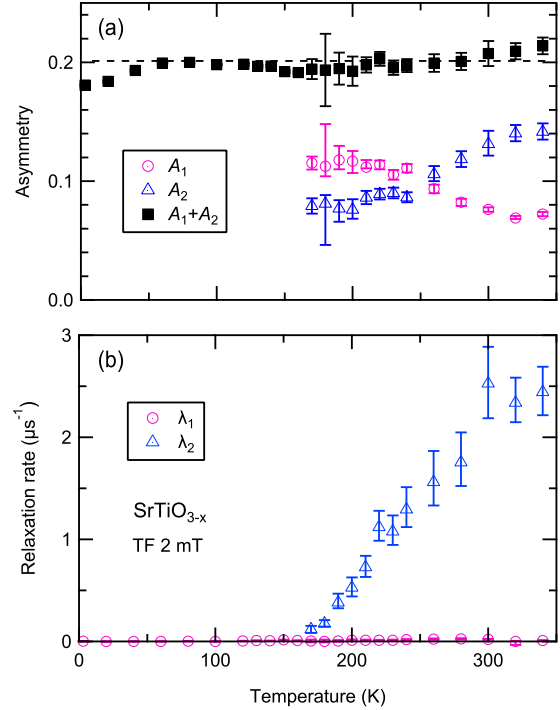}
\caption{(Color online) Temperature dependences of (a) the partial asymmetries $A_1$ and
$A_2$ and (b) the exponential relaxation rates $\lambda_1$ and
 $\lambda_2$ in SrTiO$_{3-x}$. These values were updated from those in
 Ref.~[\citen{shimomura14}] by taking account of double-pulse
 correction\cite{ito10} in a series of fits. 
 The dashed line indicates the maximum asymmetry for the
 D$\Omega$1 $\mu^+$SR spectrometer.}
 \label{fig_odef-sto_params}
\end{center}
\end{figure}

Motivated by these theoretical predictions, Shimomura {\it et~al.} 
conducted $\mu^+$SR experiments in oxygen-deficient
SrTiO$_{3-x}$ at the D1 area of J-PARC MUSE, Japan, with the expectation
that implanted muons are preferentially trapped in V$_{\rm O}$ where
they form a substitutional Mu$^-$ center.\cite{shimomura14} 
The SrTiO$_{3-x}$ sample was prepared by reducing single-crystalline wafers of
SrTiO$_3$ with the (110) surface according to the procedure outlined in 
Ref.~[\citen{jalan08}].
The $\mu^+$SR measurements were performed using a double-pulsed surface
muon beam and the D$\Omega$1 spectrometer (maximum asymmetry $\sim
$~0.20) in the LF geometry\cite{yaounc10} under a weak TF of 2~mT applied perpendicular
to the [110] direction.

$\mu^+$SR asymmetry spectra were fitted to a function composed of
two exponentially-damped cosines with a shared diamagnetic frequency.
Following [\citen{shimomura14}], the fits were refined by taking into account double-pulse
correction.\cite{ito10} 
Figure~\ref{fig_odef-sto_params} shows the partial asymmetries $A_1$ and
$A_2$ and the exponential relaxation rates $\lambda_1$ and $\lambda_2$
for the two (fast and slow) components as functions of temperature. The $A_2$ for the fast component was set to zero
for fits below 160~K.
We can find a decrease in the total diamagnetic asymmetry, $A_1+A_2$,
below $\sim$~50~K from the maximum value of $\sim$~0.20. Similar behavior 
was also observed in nominally undoped SrTiO$_3$, where muons
should stay at interstitial sites, and the asymmetry drop was 
attributed to the formation of shallow hydrogen-like Mu\cite{salman14} or Mu$^+$-bound small
polaron\cite{ito19}.
This similarity implies that an interstitial Mu$^+$ state is partially involved in the
observed diamagnetic signal. However, the diamagnetic amplitude below 50~K is
significantly larger than that in undoped SrTiO$_3$. Shimomura
{\it et al.} ascribed the increased diamagnetic fraction in
oxygen-deficient SrTiO$_{3-x}$ to the hydridic Mu$^-$ center formed in V$_{\rm O}$.

The $\lambda_2$ for the fast component markedly increases with increasing
temperature above 160~K. Shimomura {\it et al.} tentatively attributed this anomalous
behavior to the
activation of cyclic Mu$^{-/0}$ charge-state transitions on the basis of
similarity with the temperature dependence of $1/T_1$ in $n$-type GaAs\cite{chow96}
(see Section~\ref{method_thermal}).
However, there seem to be some difficulties in this interpretation.
In particular, the Mu$^{-/0}$ charge cycle model for $n$-type materials assumes that the Mu$^-$ defect creates a
deep level in the band gap and serves as a recombination
center.\cite{chow96} However, according to computational studies, the
defect level for the substitutional H$^-$ falls in the valence
band.\cite{iwazaki14}
Moreover, a characteristic energy of $\sim$~0.1~eV estimated from
the increase in $\lambda_2(T)$ is much smaller than the band-gap
energy of 3.2~eV. Accordingly, a different model may be necessary to explain the
unusual exponential relaxation possibly associated with Mu-state dynamics in SrTiO$_{3-x}$.

\subsection{Metastable Hydride-Related Center in BaTiO$_{3-x}$H$_x$
  [\citen{ito17}]}
  \label{review_btoh}
Oxyhydrides of perovskite titanates $A$TiO$_{3-x}$H$_x$ ($A={\rm Ba}$, Sr, and Ca)
comprise a new class of hydrogen ion conductors, which can be obtained from
$A$TiO$_3$ parent insulators by CaH$_2$ reduction.\cite{kobayashi12,yajima12,masuda15}
Conventional diffraction techniques can be applied to identify the H
site in $A$TiO$_{3-x}$H$_x$ since a large amount of H up to $x\sim
0.6$ can be incorporated.
A combined analysis of x-ray and neutron diffraction data revealed that 
O$^{2-}$ ions in the perovskite lattice are randomly substituted by
H$^-$ ions without creating any detectable amount of V$_{\rm O}$.
The substitutional H$^-$ in $A$TiO$_{3-x}$H$_x$ is in sharp contrast with the interstitial
protonic hydrogen (formally H$^+$) bound to an O$^{2-}$ ion, which is a
common impurity often found in $A$TiO$_3$. Macroscopic gas analysis
revealed that the hydrogen in the solid phase is mobile and exchangeable
in hydrogen gas environments above $\sim$~400$^{\circ}$C. Moreover, the 
reduction treatment changes the parent band
insulators into paramagnetic metals. These transport characteristics
suggest that $A$TiO$_{3-x}$H$_x$ compounds are suitable for
application in mixed electron/hydrogen ion conductors and hydrogen
membranes.

\begin{figure}
\begin{center}
\includegraphics[scale=0.55]{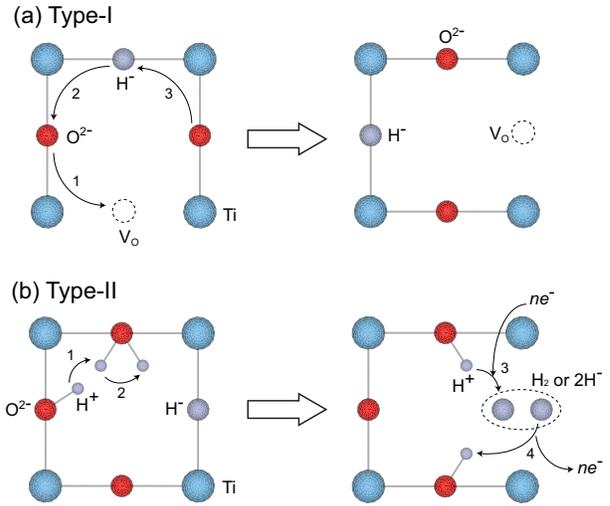}
\caption{(Color online) (a) Type-I and (b) type-II H migration models for
 $A$TiO$_{3-x}$H$_x$. The structure was drawn using VESTA.\cite{vesta3}}
\label{BTOH-migration-models}
\end{center}
\end{figure}

Several theoretical models have been proposed regarding the stability of
H species in $A$TiO$_{3-x}$H$_x$ and their
kinetics.\cite{liu17,iwazaki10,iwazaki14,zhang14} These studies
commonly conclude that the substitutional H$^-$ configuration is most stable in
$n$-type carrier-rich environments. With respect to the dynamical aspect of H in
the solid phase, two types of scenarios have been proposed.
One is based on the concept of correlated migration of H$^-$, O$^{2-}$,
and V$_{\rm O}$ in the network of the anion site [type-I H
migration, Fig.~\ref{BTOH-migration-models}(a)].\cite{kobayashi12,liu17,tang17}
The other model [type-II H migration,
Fig.~\ref{BTOH-migration-models}(b)] involves two metastable H
configurations: the interstitial H$^+$ bound to O$^{2-}$, and two H
atoms trapped at the same anion site in place of O$^{2-}$.
The former can support rapid proton diffusion via the Grotthuss
mechanism\cite{zhang14}, and the latter can act as a hydrogen exchange
center, which is expected to temporarily form due to the interaction between a substitutional
H$^-$ and an incoming interstitial H$^+$. Two charge configurations, 
hydridic 2H$^-$\cite{iwazaki14} and molecular H$_2$\cite{zhang14}, have
been proposed for this center.

The $\mu^+$SR technique is useful for creating and investigating 
experimentally accessible models of such metastable H-related centers.
This is based on the fact that the as-implanted mixture of Mu states is
far from equilibrium, often involving metastable excited states.\cite{lichti08}
Ito {\it et al.} conducted $\mu^+$SR investigations of prototypical
BaTiO$_{3-x}$H$_x$ to obtain microscopic insight into the metastable H
configurations associated with type-II H migration using Mu as a
pseudoisotope of H.\cite{ito17}
The $\mu^+$SR measurements of powder samples of BaTiO$_{3-x}$H$_x$
($x=0.1$, 0.2, 0.3 and 0.5) were carried out in the D1 area of J-PARC
MUSE, Japan, and in the port 2 of RIKEN-RAL, U.K., using a spin-polarized surface
muon beam.
Muons implanted in the samples lose their kinetic energy via 
electromagnetic interactions with host atoms and are then trapped in
local potential minima (not necessarily in the global minimum). 
In BaTiO$_{3-x}$H$_x$, implanted muons are expected to serve as
incoming excess hydrogen, creating Mu analogs of the metastable configurations
involved in the type-II H migration model together with H$^-$ in the
host lattice. Information on their charge states was obtained via the
local structure approach outlined in Section~\ref{method_structure}.

\begin{figure}
\begin{center}
\includegraphics[scale=0.58]{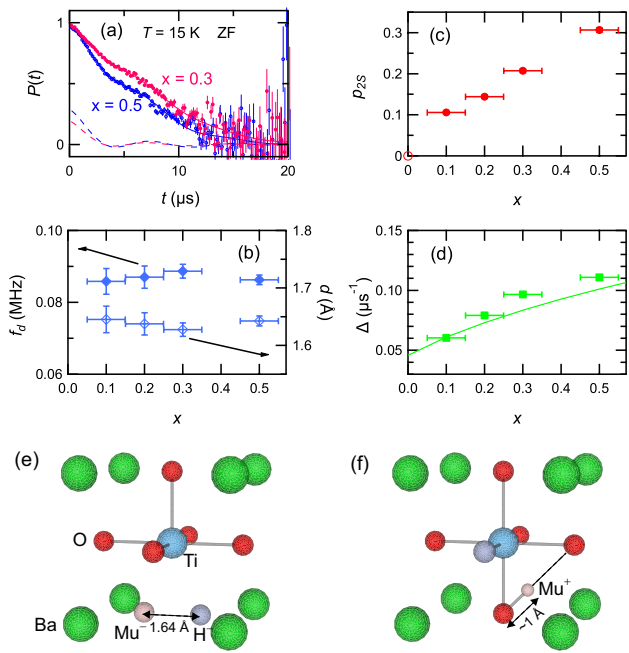}
\caption{(Color online) (a) ZF-$\mu^+$SR spectra of BaTiO$_{3-x}$H$_x$ ($x=0.3$ and
 0.5) at 15~K. The solid curves represent the best fits to
 eq.~(\ref{btoh_fit}). The dashed curves show partial contribution from
  the 2S component. (b), (c), and (d): $x$ dependences of $f_d$ and $d$,
 $p_{\rm 2S}$, and $\Delta$ at 15 K in ZF. (e) and (f): the most probable
 atomic configurations associated with the 2S and Gaussian
 components. In the structure model (f) assigned to the Gaussian component,
 O$^{2-}$ ions at the third nearest neighbor and further anion sites are
 randomly replaced by H$^-$ with a probability of $x/3$. The structure
 was drawn using VESTA.\cite{vesta3} 
 Reproduced with some modifications from Ref.~[\citen{ito17}] \copyright
 American Physical Society.}
 \label{BTOH-parameters}
\end{center}
\end{figure}

Figure~\ref{BTOH-parameters}(a) shows the time evolution of muon spin polarization $P(t)$ at
15 K in ZF for the $x=0.3$ and 0.5 samples. The $P(t)$ curves have a
damped cosine-like feature superposed on a Gaussian relaxation
curve. The oscillating feature is a signature of the formation of the
entangled 2S state composed of diamagnetic Mu and H. 
The Gaussian relaxation part is
usually ascribed to muons that interact with a large number of
surrounding nuclei without creating such a special magnetic
coupling\cite{hayano79}. Therefore, the two components were tentatively
assigned to the metastable
configurations involved in the type-II model: the 2S component
for a Mu analog of the hydrogen exchange center [Fig.~\ref{BTOH-parameters}(e)] and the
Gaussian component for the interstitial Mu$^+$ [Fig.~\ref{BTOH-parameters}(f)].

Accordingly, the ZF spectra at 15 K were fitted
to the following function, 
\begin{equation}
P(t) = p_{\rm 2S}e^{-\lambda t} G_{\rm 2S}(t;f_d)+(1-p_{\rm
 2S})e^{-\Delta^2 t^2},
 \label{btoh_fit}
\end{equation}
where $\lambda$, $p_{\rm 2S}$, and $G_{\rm 2S}(t;f_d)$ denote the relaxation rate, the
fraction, and the ZF relaxation function in eq.~(\ref{eq_2S}) for the 2S
state, respectively. The distance $d$ between a muon and a nearby H can
be obtained from $f_d$ through eq.~(\ref{eq_fd}).
$d$ is independent of $x$ [Fig.~\ref{BTOH-parameters}(b)], which is
expected since the Mu-H structure should be robust against changes in $x$.
The average value of
$d=1.64(1)$~\AA~ is consistent with the theoretical H$^-$-H$^-$ distances of 1.64
and 1.67~\AA~ for symmetric and asymmetric 2H$^-$ centers in
SrTiO$_{3-x}$\cite{iwazaki14}; however, this does not match $d=0.74$~\AA~ for a
molecular MuH configuration. On the basis of these results, Ito {\it et al.}
assigned the 2S component to a hydridic (Mu$^-$, H$^-$) complex formed at
V$_{\rm O}$. 
Meanwhile, the $x$ dependence of the Gaussian relaxation rate $\Delta$ associated
with the interstitial Mu$^+$ configuration can be reproduced by 
calculating the rms width of the nuclear dipolar field at the
interstitial Mu$^+$ site, as shown in Fig.~\ref{BTOH-parameters}(d) with a solid curve.

The temperature dependence of $\Delta$ for the
$x=0.5$ sample reveals that interstitial Mu$^+$
diffusion (activation energy $\sim$~0.1~eV) and subsequent trapping in an
unknown deep potential well occur above 100~K within the $\mu^+$SR time
window. This re-trapping process may function as a rate-limiting step of
macroscopic H transport in the BaTiO$_{3-x}$H$_x$ lattice.
It is not taken into account in the type-II H migration model.


\section{Future Prospects}
 The identification of Mu$^-$ and related defects in solids remains
 a challenge due to the fact that an $easy$ method of separating Mu$^-$ from Mu$^+$
 is yet to be established. A lot of beam time is still required to scan the vast parameter space of temperature, field, and carrier
 concentration to accumulate indirect evidence of Mu charge states 
 and secure overall consistency with theoretical predictions.
 Pulsed muon beams are beneficial for these 
 experiments due to the capability of the high-count-rate measurements.
The major drawback of the pulsed $\mu^+$SR method (i.e., relatively low
 time resolution) can be mostly overcome by using the RF resonance technique.
 The RF technique is also indispensable with respect to the
 sensitivity to the Mu final state formed via a delayed
 process.\cite{kreitzman91,hitti99,lichti99,lichti07,lichti08}
 Unfortunately, the RF-$\mu^+$SR method has only occasionally been used for materials
 research, even though the majority of muon facilities support
 it. Obviously, further improvements of RF
  instruments (and their user interfaces) are vital for future
  Mu$^-$ studies.
  
The Mu chemical shift approach reviewed in
Section~\ref{method_chemical_shift} and Section~\ref{review_mayenite} is
undoubtedly promising for separating Mu$^-$ from Mu$^+$; however, it also 
seems to require further improvements. This approach is based on 
comparisons with computational and/or experimental $^1$H chemical shifts.
Unfortunately, the use of the different standard samples (TMS for
$^1$H-NMR\cite{hayashi14} and CaCO$_3$ for $\mu^+$SR\cite{hiraishi16})
may have resulted in some ambiguity in said comparison.
These materials were employed to set a chemical shift of zero. It should
be noted that the zero shift does not necessarily mean a null
chemical-shielding effect on the $^1$H nuclei or muons. First-principles calculations
revealed that the absolute screening constant for protons in TMS is
about 30~ppm.\cite{baldridge99}. This relatively high shielding 
results in the majority of $^1$H resonances occurring downfield
within 20~ppm of TMS.\cite{akitt92} Meanwhile, little is known
about the chemical-shielding effect on the diamagnetic Mu species in CaCO$_3$. The
relation between the TMS and CaCO$_3$ references must be clarified for 
establishing the Mu chemical shift strategy with respect to separating diamagnetic
Mu species. The absolute screening constant for the diamagnetic Mu species
in CaCO$_3$ may be determined with high precision by using the
apparatus for the MuSEUM (Muonium Spectroscopy Experiment Using Microwave) project at J-PARC.\cite{tanaka18}


\section*{Acknowledgment}
Discussions and correspondence are gratefully acknowledged with
K. Nishiyama, K. Fukutani, Y. Iwazaki, S. Tsuneyuki, R. Kadono, and N. Nishida.

\bibliography{negmu}
\bibliographystyle{jpsj}


\end{document}